\documentclass[12pt]{article}

\usepackage{graphicx}
\title{Quark-Antiquark System in Ultra-Intense Magnetic Field
}
\author{  M.A.~Andreichikov\footnote{ e-mail:andreichicov@mail.ru} $^{1)2)}$ ,
~B.O.~Kerbikov\footnote{ e-mail:borisk@itep.ru} $^{1)2)}$
,~Yu.A.~Simonov\footnote{ e-mail:simonov@itep.ru} $^{1)}$
\\ $^{1)}$ State Research
Center\\Institute of Theoretical and Experimental Physics, \\
Moscow, 117218 Russia,\\
 $^{ 2)}$ Moscow Institute of Physics and Technology,\\Moscow region, 141700
 Russia}
 \date{}
\newcommand{\beq}{\begin{eqnarray}}
 \newcommand{\eeq}{\end{eqnarray}}
\newcommand{\be}{\begin{equation}}
 \newcommand{\ee}{\end{equation}}

\def\fun#1#2{\lower3.6pt\vbox{\baselineskip0pt\lineskip.9pt
\ialign{$\mathsurround=0pt#1\hfil ##\hfil$\crcr#2\crcr\sim\crcr}}}

\newcommand{{\SD}}{\rm SD}
\newcommand{{\Lc}}{\mathcal{L}}
\newcommand{{\Mc}}{\mathcal{M}}

\newcommand{\ver}{\mbox{\boldmath${\rm r}$}}
\newcommand{\vesig}{\mbox{\boldmath${\rm \sigma}$}}

\newcommand{\veP}{\mbox{\boldmath${\rm P}$}}
\newcommand{\vep}{\mbox{\boldmath${\rm p}$}}

\newcommand{\vez}{\mbox{\boldmath${\rm z}$}}
\newcommand{\veA}{\mbox{\boldmath${\rm A}$}}

\newcommand{\veL}{\mbox{\boldmath${\rm L}$}}

\newcommand{\veR}{\mbox{\boldmath${\rm R}$}}

\newcommand{\vepi}{\mbox{\boldmath${\rm \pi}$}}
\newcommand{\veta}{\mbox{\boldmath${\rm \eta}$}}
\newcommand{\veB}{\mbox{\boldmath${\rm B}$}}
\newcommand{\veH}{\mbox{\boldmath${\rm H}$}}
\newcommand{\veE}{\mbox{\boldmath${\rm E}$}}

\newcommand{\veF}{\mbox{\boldmath${\rm F}$}}

\newcommand{\lan}{\langle}
\newcommand{\ran}{\rangle}
\begin{document}
\maketitle
\begin{abstract}
We study the relativistic quark-antiquark system  embedded in  magnetic field
(MF). The  Hamiltonian containing confinement, color Coulomb and spin-spin
interaction is derived. We analytically follow the evolution of the lowest
neutral meson state as a function of  MF strength. Calculating the color
Coulomb energy $\lan V_{\rm Coul}\ran$ we have  observed the  unbounded
negative (at least in the limit of large $N_c$)  contribution at large MF which
makes the mass negative for $eB> eB^{QCD}_{\rm crit}$. We display the $\pi^0$
and $\rho^0$ masses as functions of MF in comparison with recent lattice data.

\end{abstract}
Pacs:

\section{Introduction}

During the last years we have witnessed an impressive progress of the
fundamental physics in ultra-intense magnetic field (MF) reaching the strength
up to $eB \sim 10^{18} G \sim m^2_\pi $ \cite{1}. Until  recently magnetars
\cite{2} were the only physical objects, where such, or somewhat weaker MF
could be realized. Now MF of the  above  strength and even stronger is within
reach in peripheral  heavy ion  collisions at RHIC and LHC \cite{3}. High
intensity lasers is another perspective tool to achieve MF beyond the Schwinger
limit \cite{4}. On the theoretical side  a striking progress has been achieved
along several lines. It is beyond our scope to discuss these works  or even
present a list of corresponding references. We mention only two lines of
research which have a certain overlap with our work. The first one \cite{5,6}
is the behavior of the hydrogen atom and positronium in very strong MF. The
second one \cite{7} is the conjecture of the vacuum reconstruction   due to
 vector meson  condensation in large MF. The relation between the  above studies and our
work will be clarified in what follows.

Our goal is  to study from the   first principles  the spectrum of a meson
composed of quark-antiquark embedded in MF. Use will be made of
Fock-Feynman-Schwinger representation  (see \cite{8} for review and references)
of the quark Green's function with strong (QCD) interaction and MF included. An
alternative approach could have been Bethe-Salpeter type formalism. However,
for  the confinement originating from the area law of the Wilson loop, the  use
of the gluon propagator is inadequate. Numerous attempts in this direction
failed because of gauge dependence and the vector character of the gluon
propagator, while confinement is scalar and gauge invariant. Therefore it is
sensible to use the path integral technique for QCD$+$QED Green's functions.
This method in combination with the einbein  technique (the method of effective
masses) \cite{9} enables one to construct explicit expressions for meson
Hamiltonians without MF \cite{10}. In this way spectra of light-light,
light-heavy and heavy-heavy mesons were computed with a good accuracy, using
the string tension $\sigma$, strong coupling constant $\alpha_S$ and quark
current masses as an input \cite{11},\cite{I}.

In what follows we expand this technique  to incorporate the effects of MF on
mesons. The latter contains: 1)  direct influence of MF on quark and antiquark,
and 2) the influence on gluonic fields, e.g., on $\alpha_S$, gluon propagator
and on the gluon field correlators determining the string tension $\sigma$.
However, since MF acts on charged objects, its influence on the gluonic degrees
of freedom enters only via $(N_c)^{-k}, ~~ k=1,2,...$  In what follows the
corrections of the second type will be neglected. 3) As will be discussed
 elsewhere, MF also changes quark condensate $\lan \bar q q\ran$ and quark decay
constants $f_\pi$ etc., and in this way strongly influences chiral dynamics.

The decisive step in our relativistic  formalism is the implementation of the
pseudomomentum notion and  c.m. factorization in MF, suggested in the
nonrelativistic case in \cite{12} for neutral two particle systems.

The plan of the paper is the following. Section 2 contains a brief pedagogical
reminder of how the two-body problem in MF is solved in quantum mechanics. The
central point  here is the integral of motion (``pseudomomentum'') which allows
the separation of the center of mass. Here we also  show how to diagonalize the
spin-dependent interaction. In section 3 we formulate the path integral for
quark-antiquark system with QCD$+$QED interaction. Then from Green's function
the relativistic Hamiltonian is obtained. Section 4 is devoted to the treatment
of confining and color Coulomb terms, we  demonstrate the unboundedness of
$q\bar{q}$ spectrum due to the latter.
 This phenomenon can be called the ``magnetic QCD collapse'',
  which occurs in the large $N_c$ limit: $n_f/N_c \to 0$. Here we also present the derivation of the
 eigenvalue equations for the relativistic Coulomb problem.
 In section 5 we discuss the spectrum
of the system focusing on the regime of ultra-strong MF.  Section 6 contains
the discussion of the results,  comparison with lattice calculations, drawing
further perspectives and intersections of our results with those of other
authors \cite{5,6,7}.

\section{Pseudomomentum and Wavefunction Factorization}

The total momentum of $N$ mutually interacting particles with translation
invariant interaction is a constant of motion and the center of mass motions
can be separated in Schroedinger equation. It was shown \cite{12} that a system
embedded in a constant MF also possesses a constant of motion
--``pseudomomentum''. As a result for the case  of zero total electric charge
$Q=0$ the c.m. motion  can be removed from the total Hamiltonian\footnote{The
case $Q\neq 0$ is more complicated and will be considered elsewhere.}. The
simplest example is a two-particle system with equal masses $m_1=m_2=m$ and
electric charges $e_1 = - e_2=e$. We define \be \veR = \frac{\ver_1+\ver_2}{2},
~~ \veta= \ver_1 -\ver_2, ~~ \veP = \vep_1 +\vep_2.\label{1}\ee

Straightforward calculation  in the London gauge $\veA =\frac12 (\veB\times
\ver)$ yields \be \hat H =\frac{1}{4m} \left(\veP-\frac{e}{2} \left(\veB\times
\veta\right)\right)^2 + \frac{1}{m} \left(-i\frac{\partial}{\partial
\veta}-\frac{e}{2}(\veB\times \veR)\right)^2+V(\eta).\label{2}\ee
%%%%%%%%%%%
One can verify that the following ``pseudomomentum'' operator $\veF$ commutes
with the Hamiltonian (\ref{2}) \be \hat{\veF} =\veP +\frac{e}{2}(\veB\times
\veta).\label{3}\ee This immediately leads to the following factorization of
the wave function (WF) \be \Psi (\veR,\veta) =\varphi (\veta) \exp \left\{
i\veP \veR -i\frac{e}{2}(\veB\times \veta)\veR\right\}.\label{4}\ee

For the oscillator-type potential $V(\eta)$ the problem reduces to a set of
three oscillators, two of them are in a plane perpendicular to the magnetic
field and their frequencies are degenerate, while  the third  one is connected
solely with $V(\eta)$.

Next we briefly elucidate
 the spin interaction in presence of MF. The corresponding part of
 the Hamiltonian may be written as
 \be \hat H_s =4a_{hf} (\vesig_1 \vesig_2) -\mu\veB (\vesig_1-\vesig_2),\label{5}\ee
where $e_1=-e_2=e>0$ and $\mu>0$. Diagonalization of $\hat H_s$ yields the
following four eigenvalues  e.g. for $u\bar u$ system, comprising both $\rho$
and $\pi$ levels. \be E_{1,2}^{(s)} =a_{hf}, ~~ E_{3,4}^{(s)} =\pm a_{hf}
\left( 2\sqrt{1+\left(\frac{\mu B}{a_{hf}}\right)^2}\mp 1\right),\label{6}\ee
where we assume that $\veB$ is aligned along the positive $z$-axis and
$B=|\veB|$. In a strong MF  when $\mu B
> a_{hf}$ spin-spin interaction becomes unimportant and $E_{3,4}^{(s)} \simeq \pm
2\mu B$. For the lowest level $E_4^{(s)}$ this corresponds to a configuration
$|+-\ran$ when the spin of negatively charged particle is aligned antiparallel
to $\veB$, and the spin of the positively charged one -- parallel to $\veB$.
This means that the spin (and isospin) are no more good quantum numbers and
eigenvalues (\ref{6}) correspond to the mixture of spin 1 and spin  0 states.
As a result the $q \bar q$ state will split into 4 states (two of them
coinciding $E_1^{(s)} = E_2^{(s)}$).
 Till now
we treated a nonrelativistic system, to incorporate relativistic effects we
shall exploit the path integral form of relativistic Green's functions
\cite{8}.

\section{Relativistic $q\bar q$ Green's function and effective Hamiltonian}

The starting point is the Fock-Feynmann-Schwinger (world-line) representation
of the quark Green's function \cite{8}. The role of the ``time'' parameter
along  the path $z^{(i)}_\mu(s_i)$ of the $i$-th quark is played by the
Fock-Schwinger proper time $s_i, i=1,2.$ Consider a quark with a charge $e_i$
in a gluonic field $A_\mu$ and  the electromagnetic vector potential
$A_\mu^{(e)}$, corresponding to a constant magnetic field $B_{i}$. Then the
quark propagator in the Euclidean space-time is \be S_i (x,y) = (m_i +\hat
\partial - i g \hat A - i e_i \hat A^{(e)})^{-1}_{xy} \equiv(m_i+\hat
D^{(i)})^{-1}_{xy}.\label{7}\ee

The path-integral representation for $S_i$ is \cite{8} \be S_i (x,y) = (m_i -
\hat D^{(i)})\int^\infty_0 ds_i (Dz)_{xy}e^{-K_i}\Phi_\sigma^{(i)} (x,y)\equiv
(m_i-\hat D^{(i)}) G_i (x,y), \label{8}\ee where \be K_i = m_i^2 s_i +
\frac14\int^{s_i}_0 d\tau_i
\left(\frac{dz_\mu^{(i)}}{d\tau_i}\right)^2,\label{9}\ee $$\Phi^{(i)}_\sigma
(x,y) =P_AP_F \exp \left( ig \int^x_y A_\mu dz_\mu^{(i)}+ ie_i \int^x_y
A^{(e)}_\mu dz_\mu^{(i)}\right)\times$$ \be \times \exp \left( \int^{s_i}_0
d\tau_i \sigma_{\mu\nu} (gF_{\mu\nu} + e_i B_{\mu\nu})\right).\label{10}\ee
Here $F_{\mu\nu} $ and $B_{\mu\nu}$ are correspondingly gluon and MF tensors,
$P_A, P_F$ are ordering operators,  $\sigma_{\mu\nu} = \frac{1}{4i}
(\gamma_\mu\gamma_\nu-\gamma_\nu\gamma_\mu)$. Eqs. (\ref{7}-\ref{10}) hold for
the quark, $i=1$, while  for the antiquark one should reverse the signs of
$e_i$ and $g$. In explicit form one writes \be \sigma_{\mu\nu} F_{\mu\nu} =
\left(
\begin{array}{ll} \vesig\veH&\vesig \veE\\\vesig\veE& \vesig
\veH\end{array}\right),~~ \sigma_{\mu\nu}B_{\mu\nu} = \left(
\begin{array}{ll} \vesig\veB&0\\0&\vesig\veB\end{array}\right).\label{11}\ee
Next we consider $q_1\bar q_2$ system born at the point $x$ with the current
$j_{\Gamma_1}(x) = \bar q_1 (x) \Gamma_1 q_2 (x)$ and  annihilated at the point
$y$ with the current $j_{\Gamma_2} (y)$. Here $x$ and $y$ denote the sets of
initial and final coordinates of quark and antiquark.
 Using the nonabelian Stokes theorem
and cluster expansion for the gluon field(see \cite{11} for reviews) and
leaving the MF term intact, we can write
$$ G_{q_1\bar q_2} (x,y) = \int^\infty_0 ds_1 \int^\infty_0 ds_2
(Dz^{(1)})_{xy} (Dz^{(2)})_{xy} \lan  \hat TW_\sigma (A)\ran_A\times$$ \be
\times \exp (ie_1 \int^x_y A^{(e)}_\mu dz^{(1)}_\mu -ie_2 \int^x_y A^{(e)}_\mu
dz^{(2)}_\mu +e_1\int^{s_1}_0 d\tau_1 (\vesig \veB) -e_2\int^{s_2}_0 d\tau_2
(\vesig \veB)),\label{12}\ee where \be \hat T = tr (\Gamma_1 (m_1  -\hat D_1)
\Gamma_2 (m_2 -\hat D_2)),\label{13}\ee and $\Gamma_1 = \gamma_{\mu},\ \Gamma_2
= \gamma_{\nu}$ for vector currents, while \be \lan W_\sigma (A)\ran_A = \exp
\left( -\frac{g^2}{2} \int d \pi_{\mu\nu} (1) d\pi_{\lambda\sigma} (2) \langle
F_{\mu\nu}(1)F_{\lambda\sigma}(2)  \rangle\right),\label{14}\ee where
$d\pi_{\mu\nu} \equiv ds_{\mu\nu} + \sigma_{\mu\nu}^{(1)} d{\tau_1} -
\sigma_{\mu\nu}^{(2)} d\tau_2,$ and $ds_{\mu\nu}$ is an area element of the
minimal surface, which can be constructed using straight lines, connecting the
points $z_\mu^{(1)} (t)$ and $z_\nu^{(2)} (t)$ on the paths of $q_1$ and $\bar
q_2$ at the  same time $t$ \cite{8, 10}. Then the spin-independent part of the
exponent reduces to the confinement term $V_{\rm conf}(r)$ plus color Coulomb
potential $V_{\rm Coul}$, while spin-dependent part $V_{SD}$  depends also on
proper time variables $\tau_1, \tau_2$, (see \cite{14} for derivation and
discussion). For the case of zero quark orbital momenta with the minimal
surface, discussed above, one obtains a simple answer for $\langle
W_{\sigma}(A)\rangle_{A}$, \be \langle W_{\sigma}(A) \rangle_{A} = \exp \left(
- \int^{\tau_E}_{0} d t_E \left[ \sigma|\vez^{(1)} - \vez^{(2)}| - \frac{4}{3}
\frac{\alpha_s}{|\vez^{(1)} - \vez^{(2)}|}  \right] \right),\label{15a} \ee
containing $V_{\rm conf}(|\ver|)$ and $V_{\rm Coul}(|\ver|)$. Here $\sigma$ is
the string tension, $\sigma = 0.2 \ \mathrm{GeV}^2$ in our calculations.

At this point we introduce the method of  einbein variables (effective masses)
$\omega_i$ defined via the connection between the proper time $\tau_i$ and the
real time $t_i^E = z_4 (\tau_i)$ \be d\tau_i = \frac{dt^E_i}{2\omega_i},~~ \int
ds_i (D^{(4)}z^{(i)})_{xy} = const \int D\omega_i (t) (D^{(3)}
z^{(i)})_{xy}.\label{15}\ee In this way the path integral in $Dz_4^{(i)}$ is
replaced by $D\omega^{(i)}$, and the latter
 can be denoted as: $\int D\omega^{(1)}D\omega^{(2)}[...] \equiv \langle [...] \rangle_{\omega} $, see \cite{15} for the  details.

First we need to find the Hamiltonian $H_{q_1 \bar q_2}$ of the system at
$t_1^E = t_2^E = t^E$. To this end we define the Euclidean Lagrangian $L^E_{q_1
\bar q_2} $. We write $\frac{d z^{(i)}}{d\tau_i} =
2\omega_i\frac{dz^{(i)}_k}{dt^E} = 2\omega_i \dot{z}_k,\ k = 1,2,3$. Then all
terms in the exponents in (\ref{12}), (\ref{14}) and (\ref{15}) can be
represented as $\exp(-\int dt^E L_{q_1 \bar q_2}^E)$ and thus we arrive at the
following action $$ S^E_{q_1\bar q_2} = \int^{T_E}_0 dt^E \left[
\frac{\omega_1+\omega_2}{2} +
 \sum_i \left( \frac{\omega_i}{2} (\dot z_k^{(i)})^2 \right)-\right.$$ \be- ie_i A_k^{(e)} \dot z^{(i)}_k + \frac{m_1^2}{2\omega_1} +
 \frac{m_2^2}{2\omega_2}+ e_1 \frac{\vesig_1 \veB}{2\omega_1}+e_2 \frac{\vesig_2
 \veB}{2\omega_2}
  \left.+\sigma |\vez^{(1)} -\vez^{(2)}| -\frac43
 \frac{\alpha_s}{|\vez^{(1)}-\vez^{(2)}|^2}\right].\label{16}\ee

 Here $A_k^{(e)}$ is the $k$--th
 component of the QED vector potential, $\sigma$ is the QCD string tension. The next step
 is the transition to the Minkowski metric. This is easy since confinement is already expressed
 in terms of string tension. We have $\exp(-\int L^E dt_E) \rightarrow \exp(i \int L^M dt_M),\ t_E \rightarrow it_M$,
  and \be H_{q_1\bar q_2} = \sum_i \dot z_k^{(i)} p^{(i)}_k -L_M, ~~ p^{(i)}_k =
 \frac{\partial L^M}{\partial\dot z_k^{(i)}} =\omega_i \dot z_k^{(i)} + e_i
 A_k^{(e)}.\label{17}\ee

 Next comes the key point of the Method of Effective Masses \cite{9,10}.
 It comprises the replacement of the path integral averaging over $\omega_1,\omega_2$ by the stationary
  point analysis. The applicability of this approximation may be justified by the following arguments.
  The $q \bar q$ Green's function (\ref{12}) integrated over $d^3(x-y)$ takes the ``heat--kernel''
  form \be G_{q_1 \bar q }(x,y) = \lan x | \hat T \exp (- H_{q_1\bar q_2} T ) |
y\ran_{\omega_1, \omega_2}\label{20}\ee
  Integrating (\ref{12}) over $d^3(x-y)$, one obtains a simple expression:
  \be \int G_{q_1 \bar{q}_2}^{(\mu \nu)}(x,y)d^3(x-y) = \left\langle
  \sum_{n,\lambda} \frac{\varepsilon_{\mu}^{(\lambda)}
  \varepsilon_{\nu}^{(\lambda)}(f_n^{(\lambda)})^2}{2M_n^{(\lambda)}}
   e^{-M_n^{(\lambda)}|x_4 - y_4|}  \right\rangle_{\omega_1, \omega_2} \ee
    Here  $\varepsilon_{\mu}^{(\lambda)}$ is the polarization vector for the
    polarization state $\lambda$, and $M_n^{(\lambda)}$, $f_n^{(\lambda)}$ are
     correspondingly the Hamiltonian eigenvalue and quark decay constant,
      $2M_n^{(\lambda)}$ in the denominator stems from the normalization of
       the relativistic wave functions, $n$ runs through all ordering numbers
        of the spectrum. All these quantities are functions of $\omega_1,\ \omega_2$.
         Therefore the integral (20) may be symbolically written as
         $ \langle Ke^{-MT} \rangle_{\omega_1,\omega_2} = \int D \omega_1 D
          \omega_2 K(\omega_1,\omega_2)e^{-M(\omega_1,\omega_2)T} $ and it
          is essentially defined by the region of the stationary point of the
           exponent.

The effective masses $\omega_i$ are to be found from the minimum of the total
mass $M(\omega_i)$ , as it was suggested in \cite{10}. To introduce the
minimization procedure and to check its accuracy we shall begin by the
calculation of the eigenvalues of one and two quarks in MF, and the energy of
the ground state of a relativistic charge in the atom  in the next section,
reproducing the known exact results.

           We have the following equations defining $\omega_i$
            from the total mass $M(\omega_i)$ \be \hat H \psi = M
 (\omega_i) \psi, ~~ \frac{\partial M(\omega_i)}{\partial \omega_i} =
 0.\label{19}\ee

For a single quark in MF the first of the above equations gives \be M(\omega) =
\frac{p^2_z +m^2_q+ |eB| (2n+1) - eB\sigma_z}{2\omega} +
\frac{\omega}{2}.\label{20}\ee

Then the second equation yields the correct answer \be \bar M_n = ( p^2_z+
m^2_q+ |eB| (2n+1) - eB \sigma_z)^{1/2}.\label{21}\ee
 Now we turn to the case
of $q_1 \bar q_2$ system and introduce the coordinates  which are the
generalization of (\ref{1})

\be \veR = \frac{\omega_1 \vez^{(1)} + \omega_2 \vez^{(2)}}{\omega_1
+\omega_2},~~ \veta = \vez^{(1)} - \vez^{(2)},\label{22}\ee

\be \veP=-i\frac{\partial}{\partial\veR}, ~~ \vepi =-i
\frac{\partial}{\partial\veta}.\label{23}\ee

It is convenient to introduce the following two additional parameters \beq
\tilde \omega = \frac{\omega_1 \omega_2}{\omega_1 + \omega_2},\ s =
\frac{\omega_1 - \omega_2}{\omega_1 + \omega_2} \eeq As before, for simplicity
we consider only the neutral meson, so that $e_1 = -e_2 = e$. Then the total
Hamiltonian may be written as \beq H_{q_1 \bar q_2} = H_B + H_{\sigma} + W,
\label{fh} \eeq where
$$
H_B=\frac{1}{2\omega_1}\left[ \frac{\tilde \omega}{\omega_2} \veP +\vepi
-\frac{e}{2} \veB \times \left( \veR + \frac{\tilde \omega}{\omega_1}
\veta\right) \right]^2+$$
$$  + \frac{1}{2\omega_2}\left[
\frac{\tilde \omega}{\omega_1} \veP -\vepi +\frac{e}{2} \veB \times \left( \veR
-\frac{\tilde \omega}{\omega_2} \veta\right) \right]^2 = $$ \beq
\frac{1}{2\tilde \omega}\left(\vepi -\frac{e}{2}\veB \times \veR + s
 \frac{e}{2}\veB \times \veta \right)^2 + \frac{1}{2(\omega_1 + \omega_2)}
 \left(\veP - \frac{e}{2}\veB \times \veta \right)^2.\label{a24}  \eeq
 Equation (\ref{a24}) is an obvious generalization of (\ref{2}). The two other terms in (\ref{fh}) read
\be H_\sigma = \frac{m^2_1 + \omega^2_1 - e\vesig_1 \veB}{2\omega_1} +
\frac{m_2^2 +\omega_2^2 + e \vesig_2 \veB}{2\omega_2},\label{26}\ee \be
W=V_{\rm conf} + V_{\rm Coul} + \Delta W = \sigma |\veta| -\frac43
\frac{\alpha_s(\eta)}{\eta} +  \Delta W,\label{27}\ee and $\Delta W$
 contains self--energy and spin--spin contributions. One can verify that the
  ``pseudomomentum'' operator in (\ref{3}) introduced in Section 2 commutes
  with $H_B$ and hence we can again separate the c.m. motion according to the
  ansatz (\ref{4}). Then the problem reduces to the eigenvalue problem for
  $\varphi(\veta)$ with the Hamiltonian $H_B$ having the following form:
  \beq H_B = \frac{1}{2\tilde \omega}\left(-i \frac{\partial}{\partial \veta}
  + s \frac{e}{2}\veB \times \veta  \right)^2 + \frac{1}{2(\omega_1 + \omega_2)}
  \left(\veP - e\veB \times \veta \right)^2 \eeq For $\veP \times \veB = 0$ the system
   has a rotational symmetry and the c.m. is freely moving along the $z$-axis. Here
    we shall consider a state with zero orbital momentum $(\veL_{\eta})_z =
    \left[ \veta \times \frac{\partial}{i\partial \veta}  \right]_z = 0$.
    As a result $H_B$ is replaced by a purely internal space operator
    \be H_0 = \frac{1}{2\tilde \omega} \left( - \frac{\partial^2}{\partial
     \veta^2}+\frac{e^2}{4} (\veB\times \veta)^2 
     \right),\label{28}\ee
     To test our method we put $W = 0$ and arrive at the equation
     \be (H_0 + h_\sigma) \varphi = M(\omega_1,\omega_2) \varphi.\label{29}\ee
      Consequent minimization of $M(\omega_1,\omega_2)$ in $\omega_1, \omega_2$ ,
       as in (\ref{21}) , yields the expected answer for the two independent quarks, \be M=\sqrt{m_1^2 + eB (2n_1+1) - e\vesig_1
\veB}+\sqrt{m_2^2 + eB (2n_2+1) + e\vesig_2 \veB}.\label{30}\ee

\section{Treating confinement and color Coulomb terms. The magnetic QCD collapse}

From{(\ref{27}), (\ref{28}) it is clear, that inclusion of $V_{\rm conf}$ and
$V_{\rm Coul} $ in $H_0 + W $ leads to a differential equation in variables
$\eta_\bot, \eta_z,$ which can be solved numerically. However, in order to
obtain a clear physical picture, we shall represent $V_{\rm conf}$ in a
quadratic form. This will allow to get an exact analytic solution in terms of
oscillator functions with eigenvalue  accuracy of the order of $5\%$. The color
Coulomb contribution will be estimated as an average $\lan \varphi |V_{\rm
Coul}|\varphi\ran$, thus yielding an upper limit for the total mass.

For $V_{\rm conf}$ we choose the form \be V_{\rm conf} \to \tilde V_{\rm conf}
= \frac{\sigma}{2} \left( \frac{\eta^2}{\gamma} + \gamma\right)\label{31}\ee

Here $\gamma$ is a positive variational parameter; minimizing $\tilde V_{\rm
conf}$ w.r.t. $\gamma$, one returns to  $V_{\rm conf}$. We shall determine
$M(\omega_1 \omega_2, \gamma) $  corresponding to $\tilde V_{\rm conf}$, and to
define $\gamma$ an additional condition \be \left.\frac{\partial M (\omega_1,
\omega_2, \gamma)}{\partial \gamma}\right|_{\gamma=\gamma_0} =0\label{32}\ee
will be added to (\ref{19}). As a result  $M(\omega_1^{(0)}, \omega_2^{(0)},
\gamma_0)$ will be the final answer for the mass of the system. The difference
of the exact numerical solution from that obtained with the genuine potential
$V_{\rm conf}$ does not exceed $5 \%$ . The solution of the equation $(H_0 +
\tilde V_{\rm conf}) \varphi = M(\omega_1 , \omega_2 , \gamma_0) \varphi$ for
the ground state is \be \psi(\veta) = \frac{1}{\sqrt{\pi^{3/2} r^2_\bot r_0}}
\exp \left( -\frac{\eta^2_\bot}{2r^2_\bot} -
 \frac{\eta^2_{z}}{2r^2_0}\right),\label{33}\ee where $r_\bot = \sqrt{\frac{2}{eB}} \left( 1+ \frac{4\sigma\tilde
 \omega}{\gamma e^2B^2}\right)^{-1/4},~ r_0 = \left( \frac{\gamma}{\sigma
 \tilde \omega}\right)^{1/4}$. As we shall see below, for the lowest mass eigenvalue with $eB\gg \sigma$,  one has
$r_\bot \approx \frac{1}{\sqrt{eB}},~ r_0  \approx \frac{1}{\sqrt{\sigma}}$ and
the $(q_1,\bar q_2)$ system acquires the
 form of an elongated ellipsoid. Similar quasi--one--dimensional picture was observed before for the hydrogen--like atoms
in strong MF \cite{5,6}. In such geometrical configuration $V_{\rm Coul}$
manifests itself in a peculiar way, again similar to what happens in hydrogen,
or positronium atoms. However, as we shall see now, in QCD, at least  in the
 large $N_c$ limit, the outcome is drastic.

We turn now to the color Coulomb term. As a starting point we present another
check of our approach, namely we shall obtain the ground state energy of two
relativistic particles with opposite charges without MF  interacting via the
Coulomb potential. The corresponding Hamiltonian reads $H=H_0 + H_\sigma -
\frac{\alpha}{\eta},$ then $H \phi = M\phi$, and  for $eB =0$ we have \be M=-
\frac{\tilde \omega \alpha^2}{2} + \frac{m^2_1+ \omega^2_1}{2\omega_1}
+\frac{m^2_2+ \omega^2_2}{2\omega_2} .\label{39}\ee

Minimizing in $\omega_1$ in the limit $m_2\gg m_1$ (the hydrogen atom), one
obtains \be M=m_1 \sqrt{1-\alpha^2},\label{40}\ee which coincides with the
known  eigenvalue of the Dirac equation.

 In our  $(q_1\bar q_2)$ case one can calculate the expectation value of $V_{\rm Coul} = - \frac43
\frac{\alpha_s(\eta)}{\eta} $ with the asymptotic freedom and IR
 saturation behaviour in $\vep$--space (see \cite{16} for  a short review)
\be \alpha_s (q) =
 \frac{4\pi}{\beta_0 ln \left( \frac{q^2+M^2_B}{\Lambda^2_{QCD}}\right)},
 \label{34}\ee where $M_B$ is proportional to $\sqrt{\sigma}$,  $M_B\approx 1$ GeV \cite{16}.
  With the wavefunction (\ref{33}) the average value of
$V_{\rm Coul} $ takes form
 \be \Delta M_{\rm Coul} \equiv \int V_{\rm Coul} (q) \tilde \psi^2 (q)
 \frac{d^3q}{(2\pi)^3} =- \frac{4}{3\pi} \int^\infty_0 \alpha_s (q) dq
 e^{-\frac{q^2 r^2_\bot}{4}} I\left[
 \frac{q^2(r^2_0-r^2_\bot)}{4}\right],\label{35}\ee
where $I(a^2)= \int^{+1}_{-1} dx e^{-a^2x^2}$. Estimating the integral in
(\ref{35}), for $eB \gg \sigma$, i.e. for $r_0 \gg r_\bot$ one obtains for
massless quarks \be \Delta M_{\rm Coul} \approx - \frac{16\sqrt{\pi}}{3r_0
\beta_0} lnln \frac{r^2_0}{r^2_\bot} \approx - \sqrt{\sigma}\ln\ln
\frac{eB}{\sigma}.\label{36}\ee

With   $eB$ increasing the upper bound for the $q\bar q$ mass is boundlessly
decreasing. The exact eigenvalue should lie  even lower. Surmising (as will be
confirmed in the  next section) that the contribution of the remaining  part of
Hamiltonian  to the total mass is  $M_0 \approx 2 \sqrt{\sigma}$, we can
estimate the  upper limit of $eB$ compatible with the conditions $M_0 + \Delta
M_{\rm Coul}\geq 0$ namely \be (eB)^{QCD}_{\max} \approx \sigma \exp
\left(\exp\left( \frac{3\beta_0}{8\sqrt{\pi}}\right)\right) \approx 2.5\cdot
10^{23} G\approx 2.8 \cdot 10^4 \sigma.\label{37}\ee

As we shall see, numerical calculations yield much  smaller limit: $
(eB)^{QCD}_{max} \approx  6$ GeV$^2$. We note, that this upper limit is much
smaller, than obtained in QED from the positronium collapse \cite{5}: \be
(eB)^{QED}_{\max} = \frac{m^2_e}{4} \exp \left(\frac{\pi^{3/2}}{\sqrt{\alpha}}+
2 C_E\right) \approx  10^{40} G.\label{38}\ee

One should note, that the QCD limit (\ref{37}) is unaffected by higher order
gluon loop corrections, since those contain only gluons, not sensitive to MF.
 However, the quark loop
corrections to $V_{\rm Coul}$ are growing like $|eB|$ and can possibly ensure
the necessary screening. This is in line with QED, where such corrections
produce screening and stabilization in the hydrogen case \cite{6,7}. Therefore
the magnetic QCD collapse (\ref{36}) refers only to the large $N_c$ limit, when
the quark loops contribution can be disregarded.

We shall not elaborate here more on this problem and its significance, leaving
the topic to a dedicated paper.

\section{Meson masses in magnetic field}

Our next task is to calculate analytically the mass $M_n (\omega_1, \omega_2,
\gamma)$  of a  $(q_1 \bar q_2)$ meson. We have to  solve the equation \be (H_0
+ H_\sigma +W) \Psi_n (\eta) = M_n(\omega_1, \omega_2, \gamma)\Psi_n (\eta),
\label{41}\ee where $H_0 ,H_\sigma,W$ are given in (\ref{26}-\ref{28}). The
result is \be M_n(\omega_1, \omega_2, \gamma) = \varepsilon_{n_\bot , n_z} +
\frac{m_1^2+\omega^2_1 - e\veB \vesig_1}{2\omega_1} +\frac{m_2^2+\omega^2_2 +
e\veB \vesig_2}{2\omega_2} +\lan \Delta M_{\rm Coul} \ran  +
\Delta M_{SE},\label{42}\ee where \be \varepsilon_{n_\bot, n_z} =
\frac{1}{2\tilde \omega} \left[ \sqrt{ e^2 B^2 + \frac{4\sigma\tilde
\omega}{\gamma}} (2n_\bot +1)+ \sqrt{\frac{4\sigma \tilde
\omega}{\gamma}}\left(n_z + \frac12\right)\right] + \frac{\gamma \sigma}{2},
\label{43}\ee

$\Delta M_{\rm Coul} $
 is given by  (\ref{35}), while according to  \cite{11,I}  $V_{SS}$ and  $\Delta M_{SE}$ are given by  \beq V_{SS} =
\frac{8\pi}{9} \frac{\alpha_{hf}|\varphi_n(0)|^2} {\omega_1 + \omega_2}
(\vesig_1 \vesig_2), \Delta M_{SE} =-
\frac{2\sigma}{\pi\omega_1^{(0)}}-\frac{2\sigma}{\pi\omega_2^{(0)}}.
\label{47a}\eeq

We note that both $V_{SS}$ and $\Delta M_{SE}$ are to be considered as
corrections and contain $\omega_1^{(0)}, \omega_2^{(0)}$, obtained from
minimization of the  remaining part  of the Hamiltonian. The parameter
$\gamma_0$ (see (35)) is defined from the condition \beq\left.\frac{\partial M
}{\partial \gamma}\right|_{\gamma=\gamma_0}=
 \left.\frac{\partial\varepsilon }{\partial
\gamma}\right|_{\gamma=\gamma_0} =0,\eeq

In the lowest state $|\bar u\downarrow, u \uparrow \rangle$ we have
$\vesig_1\veB =- \vesig_2 \veB =B$, $\omega_1^{(0)} = \omega^{(0)}_2 \equiv
\omega^{(0)}$, and $\omega^{(0)}$ is obtained from $\frac{\partial M}{\partial
\omega_i^{(0)}} = 0, \ i = 1,2$.

\begin{figure}[h]
  \centering
  \includegraphics[width=1.0\textwidth]{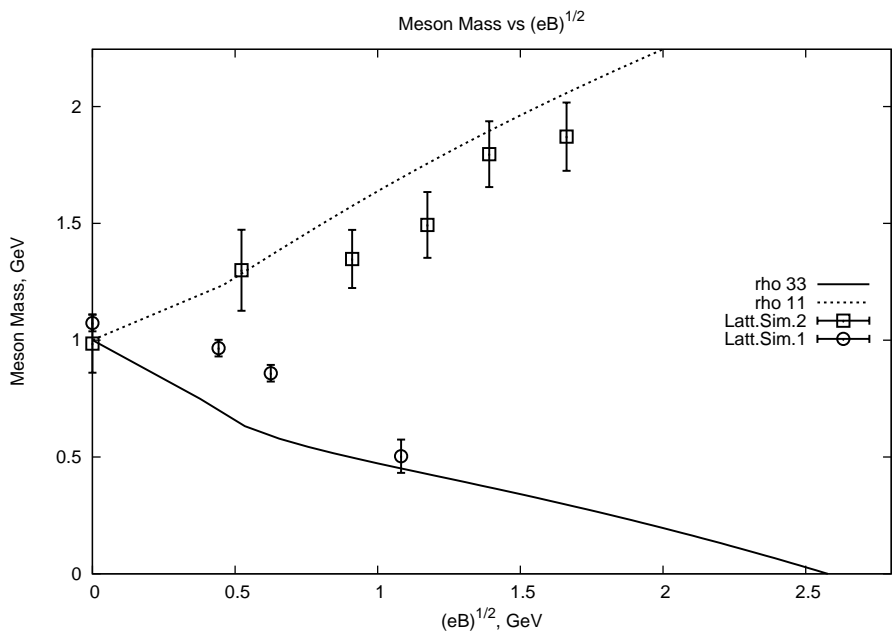}
  \caption{The mass of the system as a function of $\sqrt{eB}$. See the text for explanations.}
\end{figure}

In Fig.1 we plot the mass of the system as a function of $\sqrt{eB}$.
Calculations were performed according to (46) and the minimization procedure.
The solid curve corresponds to the configuration $|\bar u\downarrow, u \uparrow
\rangle$, the dashed one --- $|\bar u\uparrow, u \uparrow \rangle$. The circles
are from lattice calculations \cite{17}, the squares --- from \cite{18}. One
can see that the mass is slowly decreasing and reaches zero at $
(eB)^{QCD}_{Crit}$ (note, that the results plotted in Fig.1 were obtained for
massless quarks). The behaviour is in agreement with the ``magnetic QCD
collapse'' phenomena discussed above.

\section{Discussion and conclusions}

In our treatment of relativistic quark--aniquark system embedded in MF we
relied on pseudomomentum factorization of the wave function and relativistic
einbein technique. The Hamiltonian for neutral mesons in MF, containing
confinement, colour Coulomb and spin interaction was derived. Using a suitable
approximation for confining force we were able to calculate analytically meson
masses as functions of the MF. Our eye was predominately on the lowest level
with its mass decreasing with MF growing. This state is a mixture of the
$\rho^0$ and $\pi^0$ as can be seen from its spin and isospin structure.
Indeed, $u \bar u$ system under consideration is a mixture of isospin $I = 0$
and $I=1$ states, and at large MF it has a spin structure $ |u \uparrow, \bar u
\downarrow \rangle$, which is a mixture of $S=0$ and $S=1$ states. With MF
growing the mass of this state decreases (see Fig.
%\ref{fig2}
1), while the masses of all other states increase as $\sqrt{eB}$. A significant
point is that these results are in line with recent lattice simulations
\cite{17,18} (see Fig.1). Calculating the Coulomb energy $\langle \Delta M_{\rm
Coul} \rangle$ we have obtained the unbounded negative contribution at large MF
proportional to $\left( -\sqrt{\sigma}\ln \ln \frac{eB}{\sigma} \right)$ which
makes the total mass negative for $ eB > eB^{QCD}_{crit} \simeq 10 \
\mathrm{GeV}^2$.

Unlike the situation in hydrogen--like atoms, where loop corrections are able
to produce saturation \cite{6,7}, in QCD gluon loops are MF blind and uncapable
to improve the results, while quark loops are suppressed in large $N_c$ limit.
We call this problem ``magnetic collapse in QCD'' and plan to discuss it in
detail in a separate paper where quark loops and gluon polarization operator
will be considered, and possibly improve the situation, similarly to the
hydrogene atom case.

In this paper to simplify things we started with $\rho^0$ meson states
\footnote{In fact $u \bar u$ is a mock $\rho^0$} at $B = 0$ taking $\gamma_i$
in place of $\Gamma_1$ and $\Gamma_2$ in (\ref{13}). In this way we essentially
left aside the complicated problem of chiral dynamics and pseudo--Goldstone
spectrum. As explained above in this oversimplified picture we can consider the
lowest state as a mixture of $\rho^0$ and $\pi^0$. These two states are
splitted by hyperfine interaction,  this splitting is insignificant in the
large MF limit.  It is legitimate to compare the results of such treatment with
the lattice data \cite{17,18} since in the latter the quark masses are not
small and thus the chiral facet of $\pi^0$ is suppressed. As shown in Fig.1 our
analytical results are in agreement with lattice calculations \cite{17,18} both
for $\rho^o(\bar u u)$ states $|\uparrow\downarrow>$ and
$|\uparrow\uparrow>$. The behavior of the total mass $M_0$ supports the
conjecture of the ``magnetic QCD collapse'' existing in absence of quark loop
corrections.

The methods used above can be generalized to the charged states thus shading
the new light on the problem of charged  vector boson condensation suggested in
\cite{7}. As a preliminary foresight we note, that instability is an inherent
property of elementary spin 1 bosons, while $\rho$--meson can not be considered
as such an object, when MF is so strong that the Larmour radius is equal or
smaller than its size. Another system which can be treated using the same
technique is the neutral 3--body system, like neutron. The results might be
important for the neutron stars physics.

The authors are grateful for useful discussions with V.A.Novikov,
M.I.Vy\-sotsky and S.I.Godunov. We  are  indebted  to  V.S.Popov  for important
remarks.  B.K.  gratefully  acknowledge  the  support  RFBR  grant
10-02093111-NTSNIL-a. We are pleased to thank M.Chernodub for his remarks and
suggestions in response to v.1 of this paper.

\end{document}